\documentclass[11pt]{article}

\usepackage[letterpaper,margin=1in]{geometry}
\usepackage{microtype}
\usepackage{amsmath,amssymb}
\usepackage{booktabs}
\usepackage{siunitx}
\usepackage{graphicx}
\usepackage{subcaption}
\usepackage{longtable}
\usepackage{array}
\usepackage{setspace}
\usepackage{enumitem}
\usepackage[colorlinks=true,allcolors=blue]{hyperref}
\usepackage{float}
\usepackage[english]{babel}
\usepackage{chngcntr}
\usepackage{tikz}
\usetikzlibrary{positioning,arrows.meta,shapes.geometric,calc,backgrounds}
\usepackage{xcolor}

\graphicspath{{./}{./figures/}{./si_figures/}{./outputs_eviction_project1_corrected/}}
\setlist[itemize]{leftmargin=*,itemsep=0.25em,topsep=0.25em}
\setlist[enumerate]{leftmargin=*,itemsep=0.25em,topsep=0.25em}
\newcommand{\secref}[1]{\hyperref[#1]{Section~\ref*{#1}}}
\newcommand{\subsecref}[1]{\hyperref[#1]{Section~\ref*{#1}}}
\newcommand{\figref}[1]{\hyperref[#1]{Figure~\ref*{#1}}}
\newcommand{\tabref}[1]{\hyperref[#1]{Table~\ref*{#1}}}
\title{Legal Infrastructure Organizes Eviction: Evidence from Philadelphia}
\author{
  Marios Papamichalis\thanks{Human Nature Lab, Yale University, New Haven, CT 06511, \texttt{marios.papamichalis@yale.edu}}
  \and
  Regina Ruane\thanks{Department of Statistics and Data Science, The Wharton School, University of Pennsylvania, 3733 Spruce Street, Philadelphia, PA 19104-6340, \texttt{ruanej@wharton.upenn.edu}}
}
\date{}

\begin{document}
\maketitle

\begin{abstract}
We analyze the filing-side legal infrastructure of eviction using 755{,}004 Philadelphia Municipal Court landlord--tenant records filed between 1969 and 2022, of which 747{,}125 are residential. Eviction in Philadelphia is organized upstream by a concentrated plaintiff-side bar, durable plaintiff--attorney dependence, repeated use of the same properties, and recurring tenant-name exposure. Between 1983 and 2022, the ten most active plaintiff attorneys handled 82.2\% of represented plaintiff-side cases per year on average, compared with 14.8\% for the ten most active plaintiffs. Large plaintiffs depend heavily on a single attorney: among plaintiffs filing at least 101 cases, 78.3\% of each plaintiff's filings are handled by that plaintiff's most-used attorney, on average. Repetition is likewise central to the docket. Across the residential filing universe, 48.8\% of cases occur at addresses with a prior filing in the preceding year, and 23.6\% at addresses with six or more prior filings; these repeats are usually filed by the same plaintiff and follow a more default-heavy, less agreement-heavy pathway. We further examine a narrower mechanism: strict switches into specialist plaintiff-side counsel, defined as a plaintiff changing attorney to one in the prior-year top ten. Filing counts rise around the switch with non-flat pre-trends, indicating organizational reconfiguration rather than a clean exogenous shock. Within-plaintiff and within-plaintiff--property comparisons yield more stable estimates: judgment by agreement, fee share, waiver language, and corrected lockout-trigger language decline, while deadline language rises. We interpret eviction as a layered upstream process in which concentrated counsel, repeated places, and recurring tenants produce filings before any courtroom bargaining or adjudication occurs.
\end{abstract}
\section*{Significance Statement}
Eviction is typically measured as a filing count, but filings emerge from institutional pathways. We isolate that upstream pathway: which plaintiffs and plaintiff attorneys repeatedly use the court, where filings recur, whether repetition at an address reflects the same plaintiff's activity, how tenant names recur across the docket, and what changes when plaintiffs strictly switch into specialist plaintiff-side counsel. Concentrated plaintiff-side legal intermediation and repeated filing at the same addresses constitute the infrastructure through which eviction is scaled before any case is bargained, written, or adjudicated.

\section{Introduction}\label{sec:introduction}

How is eviction organized before and at the moment of filing? A court filing is not a landlord's isolated decision to sue a tenant. It is an institutional act routed through legal intermediaries, repeated property relationships, and recurring use of the same court machinery. We therefore ask five upstream questions. Is eviction more concentrated among plaintiffs or among the attorneys who represent them? Do high-volume plaintiffs depend on a small number of lawyers? Are filings reproduced at the same addresses by the same plaintiffs, and do repeat-address cases move through a different procedural path? Do repeat tenant names reappear in a single landlord relationship, or circulate across addresses and plaintiffs? When plaintiffs strictly switch into specialist plaintiff-side counsel, does filing behavior or conditional case processing change?

The answer is that eviction in Philadelphia is organized through plaintiff-side legal infrastructure. The plaintiff-side bar is substantially more concentrated than the plaintiff population, and large plaintiffs route most filings through a dominant attorney. Repeated filing is not a marginal tail of the docket: nearly half of residential filings occur at an address with a prior filing in the previous year. Those repeat-address cases are usually same-plaintiff repeats and are more likely to end in default and less likely to end in judgment by agreement. Tenant recurrence has a different structure: repeat observed tenant names often reappear at new addresses and under different plaintiffs, indicating system-level docket exposure rather than a single repeated landlord--tenant dyad.

We also estimate a narrower mechanism, kept analytically distinct from the descriptive architecture: strict switching into specialist plaintiff-side counsel. Filing counts rise around the switch, but the timing of the rise---including a non-flat pre-switch trend---is consistent with organizational reconfiguration rather than a clean exogenous shock. The more stable evidence comes from within-plaintiff and within-plaintiff--property comparisons after a case is filed. Judgment by agreement, fee share, waiver language, and corrected lockout-trigger language fall, while deadline language rises. Specialist counsel reorganizes how filed cases are bargained and written; it does not uniformly intensify every sanction.

The remainder of the paper proceeds as follows. Section~\ref{sec:literature} reviews the relevant literatures. Section~\ref{sec:data} defines the Philadelphia docket, module-specific windows, and empirical designs. Section~\ref{sec:results} reports plaintiff-side concentration, repeat filing at addresses, tenant recurrence, specialist-counsel adoption, and attorney-style convergence. Section~\ref{sec:discussion} discusses implications and limitations.

\section{Literature}\label{sec:literature}

The starting point is the now well-established observation that eviction is not an isolated response to individual default, but a patterned institution of low-income rental markets. Foundational research links eviction to poverty, residential instability, unequal housing access, discrimination, and downstream health and family harms \cite{hartman2003evictions,desmond2012eviction,desmond2015eviction,desmond2017evicted,desmond2017gets,desmond2015forced,greenberg2016discrimination}. Large administrative-record projects sharpen that account by showing that formal eviction is large in scale but must be measured carefully across filings, judgments, writs, and removals \cite{gromis2022estimating,graetz2023comprehensive,collinson2024eviction,nelson2021evictions,porton2021inaccuracies}. Because many displacements occur outside the formal docket, informal eviction research and pandemic-era policy studies further motivate separating filing activity from downstream enforcement and from broader forced mobility \cite{zainulbhai2022informal,hepburn2023protecting,benfer2023covid,summers2025pathways}.

A second body of work explains why court records should be read as records of legal processing rather than transparent measures of underlying conflict. Disputes emerge through naming, blaming, claiming, bargaining, and institutional filtering; settlements are structured legal products rather than residual non-events \cite{felstiner2017emergence,mnookin1978bargaining,priest1984selection}. Repeat-player theory predicts that high-frequency actors can convert experience, specialization, and procedural familiarity into practical advantage \cite{galanter1974haves}. Housing-court studies show that these advantages operate in courtrooms marked by lawyerlessness, compressed negotiation, default, and routinized settlement \cite{shanahan2022judges,bezdek1991silence,engler2010connecting,sabbeth2022eviction,sudeall2021praxis}. Studies of counsel effects show that representation can matter, while randomized and quasi-experimental evidence cautions that the size and direction of effects vary across courts, assistance models, litigant populations, and procedural burdens such as travel and access costs \cite{seron2001impact,ellen2021lawyers,cassidy2023effects,summers2022eviction,a2023longer}.

The main theoretical object of this analysis is legal infrastructure: the organized layer of attorneys, forms, routines, and repeat relationships through which legal action becomes scalable. Work on professional eviction practice and assembly-line plaintiffs shows that plaintiff-side counsel can become a market-making intermediary rather than a passive representative of landlords \cite{wilf2021assembly,aizman2025shadow}. Socio-legal research on the bar shows that lawyers are organized into segmented professional worlds, and scholarship on legal expertise emphasizes procedural, relational, and institutional knowledge rather than doctrinal knowledge alone \cite{rabin1983revolution,sandefur2015elements,pistor2019code}. Recent civil-court work similarly treats courts as institutional sites where social need, procedure, and governance meet imperfectly \cite{shanahan2022judges}. The frame is applied here upstream in the eviction process: filing is examined as the output of concentrated plaintiff-side legal intermediaries, repeated plaintiffs, repeated properties, and specialist counsel.

The repeat-filing part of the paper is also connected to research on serial eviction, landlord strategy, racialized housing markets, and place-based legal inequality. Serial-filing scholarship shows that many landlords use the court repeatedly as a rent-collection and discipline technology, not only as a one-time route to physical removal \cite{garboden2019serial,leung2021serial,immergluck2020evictions}. Studies of racialized eviction patterns, landlord concentration, neighborhood inequality, and civil legal extraction show that filings, fees, and court debt are embedded in broader systems of property, race, and household instability \cite{hepburn2020racial,gomory2023racially,ajayi2026landlord,brito2022racial,harris2010drawing,holland2011one,leibowitz2010repairing,rutan2021concentrated,summers2023civil,summers2026settlements}. The repeated-address and tenant-recurrence modules developed below contribute to that literature by separating repeat property use from repeat tenant exposure and by asking whether repeated filing is primarily same-plaintiff, diffuse-neighborhood, or household-recycling activity.

Finally, documents, settlement text, and record consequences are treated here as background literatures rather than as primary outcomes. Work on pleadings, notices, boilerplate, tenant screening, eviction records, and legal-document quality shows that paperwork and docket traces can shape rights, bargaining power, and future access to housing \cite{summers2024evicted,humphries2019does,kleysteuber2006tenant,reosti2020we,eisenberg2024record,brantley2025record}. The present analysis maintains a narrower boundary: it studies the plaintiff-side filing infrastructure and the adoption of specialist plaintiff-side counsel.

\section{Data, measurement windows, and empirical strategy}\label{sec:data}

\subsection{Data}\label{sec:data_data}

The analysis uses 755{,}004 Philadelphia Municipal Court landlord--tenant case records filed from 1969 through 2022. The residential filing universe contains 747{,}125 cases. The record includes filing dates, plaintiffs, tenants, addresses, attorney representation fields, procedural outcomes, monetary fields, agreement text, writ issuance, and service of the alias writ of possession. Only the fields required for the plaintiff-side filing-infrastructure analysis are used here. Courtroom-specific modules---cross-side attorney pairs, judge sorting, tenant-attorney heterogeneity, settlement-template governance, and fee-driver models---are outside the scope of this study.

The analysis uses three sample windows. Concentration and plaintiff--attorney dependence rely on 1983--2022 records in which plaintiff-attorney identifiers are reliably observed. Repeat-address and tenant-recurrence analyses use the full residential filing universe from 1969--2022, since filing dates, parties, tenants, and addresses are present throughout the archive. The specialist-counsel module uses named plaintiff-attorney cases from 1983--2022, restricted to plaintiffs that strictly switch into lagged top-10 counsel; the strongest case-level specification further conditions on the plaintiff--property unit.

\subsection{Measurement windows}\label{sec:data_windows}

\begin{table}[H]
\centering
\caption{Measurement windows used in the filing-infrastructure paper.}
\label{tab:windows}
\small
\begin{tabular}{p{0.28\linewidth}p{0.20\linewidth}p{0.42\linewidth}}
\toprule
Module & Primary window & Rationale \\
\midrule
Residential filing universe & 1969--2022 & Filing dates, plaintiffs, tenants, and addresses are observed across the archive. \\
Plaintiff vs. plaintiff-attorney concentration & 1983--2022 & Plaintiff-side attorney names become sufficiently informative for concentration measurement. \\
Plaintiff dependence, attorney entry, and shared-counsel clustering & 1983--2022 & Requires stable linkage between plaintiffs and plaintiff-side attorneys. \\
Repeat-address filing, churn, and tenant recurrence & 1969--2022 & Requires filing date, address, plaintiff, and tenant identifiers rather than courtroom actor fields. \\
Strict specialist-counsel switching & 1983--2022 & Requires lagged plaintiff-attorney volume, named attorney cases, and plaintiff switching over time. \\
\bottomrule
\end{tabular}
\end{table}

\subsection{Core measures}\label{sec:data_measures}

Plaintiff and attorney concentration are summarized annually using the Herfindahl--Hirschman Index,
\[
HHI_y=\sum_e s^2_{ey},
\]
and the top-10 filing share,
\[
Top10Share_y=\sum_{e\in T10(y)} s_{ey},
\]
where \(s_{ey}\) is entity \(e\)'s filing share in year \(y\). Plaintiff dependence on counsel is measured with plaintiff-specific top-1 attorney shares, counts of unique attorneys, top-attorney entry, plaintiff filing persistence, and shared-counsel clustering. Multi-attorney cases are also audited by exploding all listed attorneys and using fractional-credit, any-top-10, and case-attorney-pair denominators.

Repeat-address filing is measured by counting prior filings at the same cleaned address during the preceding 365 days and assigning cases to churn buckets \(\{0,1,2,3\text{--}5,6+\}\). Same-day filings are excluded from prior counts. Address keys use cleaned street addresses or valid latitude--longitude pairs; ZIP- and census-tract-based fallbacks are excluded to prevent unrelated cases from being collapsed into a single ``address.'' Same-plaintiff and other-plaintiff repeats are distinguished by plaintiff identifier. Tenant recurrence is measured from name-based docket identifiers after filtering generic aliases (``tenant,'' ``occupant,'' ``John Doe''); we interpret it as repeated observed tenant-name exposure, not verified person-level recurrence.

The specialist-counsel treatment is a strict switch, defined by three joint conditions: (i) the plaintiff changes attorney, (ii) the new attorney ranks in the top ten of represented plaintiff-side filing volume in the prior year, and (iii) the plaintiff has no earlier top-10 exposure. A separate placebo marks cases in which the plaintiff's existing attorney crosses the top-10 threshold without an attorney change, which isolates threshold drift from attorney choice.

\subsection{Empirical strategy}\label{sec:data_strategy}

Most analyses are descriptive and characterize the organization of plaintiff-side filing rather than identify causal effects. We estimate binary outcomes with linear probability models and continuous outcomes with OLS, summarize concentration through annual HHI and top-share measures, and fit repeat-address regressions with controls for amount sought, representation, housing status, ZIP fixed effects, and year fixed effects.

For filing counts we add a staggered-adoption design. Let \(G_p\) denote the first month in which plaintiff \(p\) strictly switches into lagged top-10 counsel as defined in Section~\ref{sec:data_measures}; plaintiffs that never strictly switch have \(G_p=\infty\). We estimate Callaway--Sant'Anna difference-in-differences (CSDID) group-time effects \cite{callaway2021difference}, which avoid the negative-weighting bias of staggered two-way fixed effects under heterogeneous treatment timing \cite{de2020two,sun2021estimating,goodman2021difference}. For cohort \(g\) and month \(t\), the group-time contrast is
\[
ATT(g,t)=E\{Y_{pt}-Y_{p,g-1}\mid G_p=g\}
       -E\{Y_{pt}-Y_{p,g-1}\mid G_p>t \text{ or } G_p=\infty\},
\]
where \(Y_{pt}\) is plaintiff \(p\)'s monthly filing count and \(g-1\) is the cohort-specific pre-period baseline. Event-time effects are aggregated as
\[
ATT(k)=\sum_g w_{gk} ATT(g,g+k),\qquad
w_{gk}=\frac{n_g}{\sum_{g':g'+k\in\mathcal{T}} n_{g'}},
\]
where \(k=t-g\), \(n_g\) is the number of treated plaintiffs in cohort \(g\), and controls are restricted to not-yet-treated or never-treated plaintiffs. Inference uses plaintiff-cluster influence functions and a Rademacher multiplier bootstrap over \(ATT(g,t)\) cells. A supplementary quarterly run with the canonical \texttt{differences} package implementation \cite{callaway2021difference} serves as a software cross-check; it is not used to make additional causal claims.

CSDID is restricted to the filing-count timing diagnostic. Case-level binary outcomes---judgment by agreement, fee share, served writ, and agreement-clause indicators---are sparse on plaintiff-month panels and do not aggregate stably to monthly shares. We estimate them with linear probability models that include within-plaintiff and within-plaintiff--property fixed effects:
\[
Y_{ipt}=\alpha_p+\lambda_t+\beta\,\textit{StrictTop10}_{ipt}+X_{ipt}\gamma+\varepsilon_{ipt},
\]
where the strongest specification replaces \(\alpha_p\) with plaintiff--property fixed effects. These specifications absorb time-invariant plaintiff and property heterogeneity but do not render attorney choice exogenous; we therefore interpret \(\hat\beta\) as within-unit mechanism evidence.

\section{Results}\label{sec:results}

We organize the results around the upstream pipeline shown in \figref{fig:infrastructure_overview}: concentration and dependence on counsel (\secref{sec:results_concentration}), recurring properties and tenants (\secref{sec:results_repeat}), and the specialist-counsel mechanism (\secref{sec:results_specialist}--\ref{sec:results_convergence}).

\begin{figure}[H]
\centering
\begin{tikzpicture}[
    >={Stealth[length=2.5mm]},
    layer/.style={
        rectangle, rounded corners=3pt,
        draw=black!70, line width=0.6pt,
        minimum width=11cm, minimum height=1.55cm,
        align=center, font=\small,
        inner sep=4pt
    },
    arr/.style={->, line width=0.9pt, draw=black!75},
    stat/.style={font=\footnotesize\itshape, color=black!65},
    headline/.style={font=\small\bfseries}
]

% --- Layer 1: concentrated bar ---
\node[layer, fill=blue!10] (bar) {%
    \textbf{1.\ Concentrated plaintiff-side bar}\\[1pt]
    \footnotesize Top-10 attorneys handle 82.2\% of represented cases vs.\ 14.8\% for the top-10 plaintiffs
};

% --- Layer 2: dependence ---
\node[layer, fill=blue!15, below=0.55cm of bar] (dep) {%
    \textbf{2.\ Plaintiff dependence on dominant counsel}\\[1pt]
    \footnotesize Plaintiffs with $\geq$101 cases route 78.3\% of filings through one attorney
};

% --- Layer 3: repeat addresses ---
\node[layer, fill=orange!12, below=0.55cm of dep] (addr) {%
    \textbf{3.\ Recurring properties}\\[1pt]
    \footnotesize 48.8\% of filings at addresses with a prior filing within 1 yr; 23.6\% with $\geq$6 priors\\
    \footnotesize 77.9\% of repeats are same-plaintiff $\;\bullet\;$ default rises 46$\to$58\%, JBA falls 32$\to$23\%
};

% --- Layer 4: tenant recurrence ---
\node[layer, fill=orange!18, below=0.55cm of addr] (ten) {%
    \textbf{4.\ Recurring tenant exposure}\\[1pt]
    \footnotesize 31.7\% of cases involve a repeat tenant name; 60.7\% reappear at a new address,\\
    \footnotesize 69.6\% under a different plaintiff $\;\Rightarrow\;$ system-level, not single-dyad recurrence
};

% --- Layer 5: specialist counsel ---
\node[layer, fill=green!12, below=0.55cm of ten] (spec) {%
    \textbf{5.\ Strict switches into specialist counsel} \scriptsize(\textit{N}=2{,}091 plaintiffs)\\[1pt]
    \footnotesize Within-plaintiff reorganization of agreement language:\\
    \footnotesize $\uparrow$ deadline $+$10.2 pp $\;\bullet\;$ $\uparrow$ move-out $+$2.9 pp $\;\bullet\;$ $\downarrow$ waiver $-$10.9 pp\\
    \footnotesize $\downarrow$ lockout-trigger $-$5.3 pp $\;\bullet\;$ $\downarrow$ JBA $-$4.8 pp $\;\bullet\;$ $\downarrow$ fee share $-$0.6 pp
};

% --- Outcome box ---
\node[layer, fill=black!8, draw=black!70, below=0.7cm of spec,
      minimum height=1.0cm] (out) {%
    \textbf{Filing}\\[-1pt]
    \footnotesize (bargaining, adjudication, and enforcement begin only here)
};

% --- Arrows ---
\draw[arr] (bar) -- (dep);
\draw[arr] (dep) -- (addr);
\draw[arr] (addr) -- (ten);
\draw[arr] (ten) -- (spec);
\draw[arr, line width=1.1pt] (spec) -- (out);

% --- Side bracket label ---
\node[rotate=90, anchor=south, font=\small\bfseries, color=black!60]
    at ($(bar.west)!0.5!(spec.west) + (-0.55cm,0)$) {Upstream legal infrastructure};

\end{tikzpicture}
\caption{The upstream legal infrastructure of eviction in Philadelphia. Filings emerge from a layered pipeline: a concentrated plaintiff-side bar (1), plaintiff dependence on dominant counsel (2), recurring properties (3), and recurring tenant-name exposure (4). Strict switches into specialist plaintiff-side counsel (5) reorganize agreement language within plaintiffs, before any case is bargained or adjudicated. Headline magnitudes are from Sections~\ref{sec:results_concentration}--\ref{sec:results_specialist} and corresponding tables.}
\label{fig:infrastructure_overview}
\end{figure}

\subsection{Plaintiff-side concentration and plaintiff dependence on counsel}\label{sec:results_concentration}

\begin{figure}[H]
    \centering
    \begin{subfigure}[b]{0.48\linewidth}
        \includegraphics[width=\linewidth]{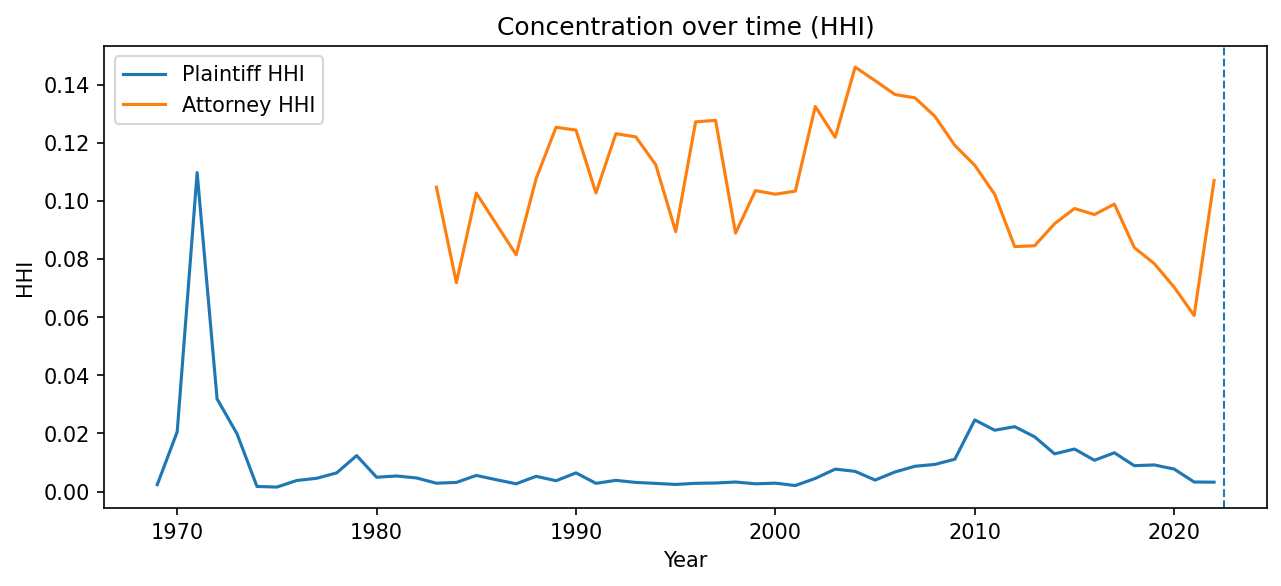}
        \caption{Annual concentration for plaintiffs and plaintiff attorneys.}
    \end{subfigure}
    \hfill
    \begin{subfigure}[b]{0.48\linewidth}
        \includegraphics[width=\linewidth]{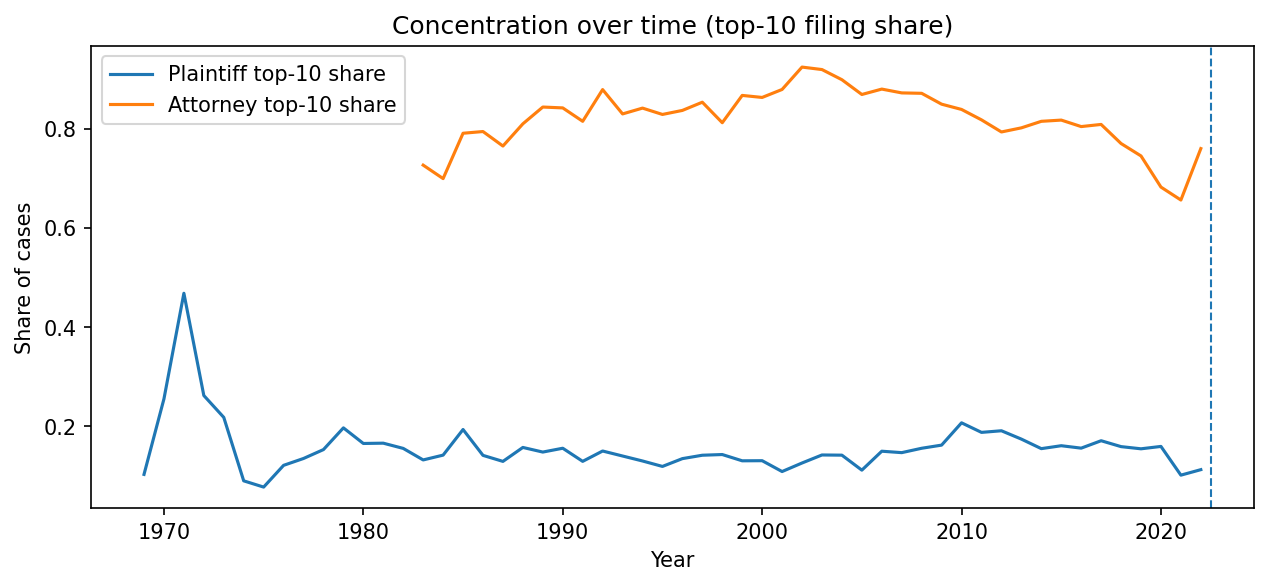}
        \caption{Annual top-10 filing shares.}
    \end{subfigure}
    \caption{Plaintiff-side concentration. The core comparison is plaintiff attorneys versus plaintiffs; the represented-field panel is used as a denominator benchmark.}
    \label{fig:legal_concentration}
\end{figure}

Concentration is much higher among plaintiff attorneys than among plaintiffs. Across 1983--2022, the ten most active plaintiff attorneys handled 82.2\% of represented plaintiff-side cases per year on average, compared with 14.8\% for the ten most active plaintiffs (mean HHI 0.107 vs.\ 0.0076). The gap is not an artifact of denominator alignment: when plaintiffs are recomputed on the represented and named-attorney universe, their mean top-10 share rises only to 0.207. The plaintiff-side bar functions as a distinct filing bottleneck (\figref{fig:legal_concentration}).

Top-10 attorneys also span many plaintiffs and many properties. Without a volume control, each top-10 attorney-year is associated with 297.8 additional plaintiffs (SE 62.8) and 532.5 additional addresses (SE 88.8), both significant at \(p<0.001\). With a log-volume control the breadth coefficients shrink but remain large (210.9 plaintiffs and 377.7 addresses, \(p<0.001\)), and top-10 status is associated with a 0.637 higher top-1 plaintiff share (\(p<0.001\)). A linear case-volume control yields a weaker portfolio interpretation. The bridge structure is therefore best read as a scale phenomenon: top-10 attorneys reach many plaintiffs and places because they operate at much larger filing volume, while their relationship to single-client dominance is nonlinear in volume.

Plaintiffs depend heavily on recurring counsel. Among plaintiffs filing at least 101 cases, 78.3\% of each plaintiff's filings are handled by that plaintiff's most-used attorney, on average; these plaintiffs use 4.17 distinct attorneys and are active for a mean of 13.5 years (Table~\ref{tab:si_dominant_counsel}). Smaller plaintiffs concentrate even more heavily on a single attorney (85.5\% in the 2--5 case bin), but they do so mechanically, because they have few cases to distribute. The substantive claim is that even the largest plaintiffs, with hundreds of cases and access to multiple attorneys, route the majority of their filings through one. Plaintiffs represented by the same attorney also have more similar plaintiff-year profiles than plaintiffs represented by different attorneys: the pooled same-minus-different distance is \(-0.129\) (permutation \(p<0.001\)). The attorney-concentration estimate is robust to multi-attorney records, with alternative any-attorney denominators producing top-10 shares between 0.813 and 0.824 (cf.\ first-listed estimate of 0.822).

\begin{figure}[H]
    \centering
    \includegraphics[width=0.70\linewidth]{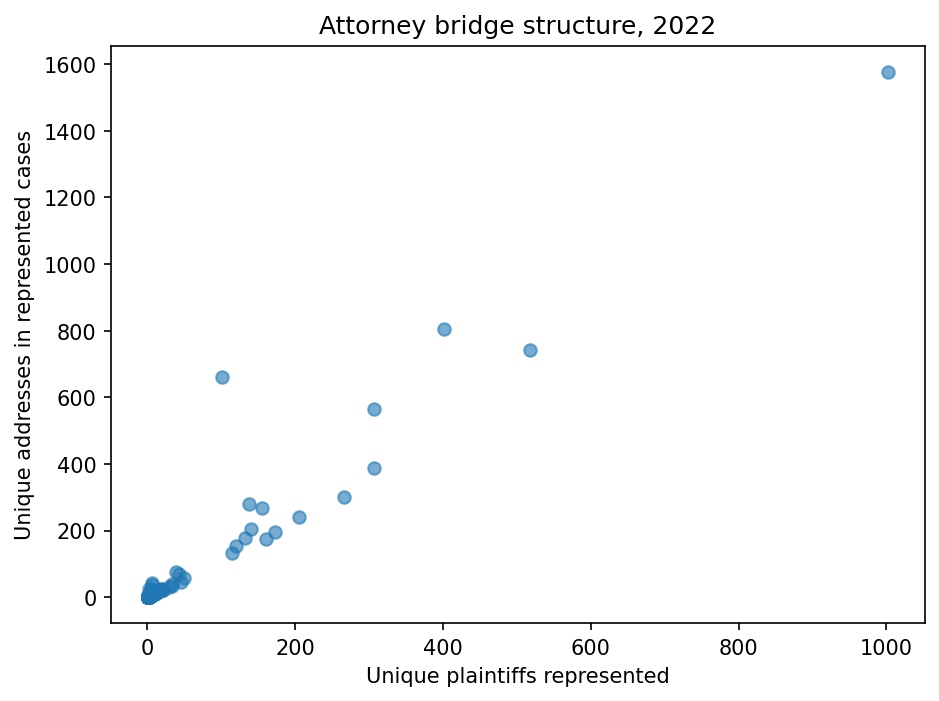}
    \caption{Attorney-year bridge structure across plaintiffs and addresses. High-volume plaintiff-side attorneys span many plaintiffs and many properties, but the relationship between top-10 status and dominant-client concentration depends on the volume specification.}
    \label{fig:legal_dependence}
\end{figure}

\begin{table}[H]
\centering
\caption{Plaintiff-side legal infrastructure: selected concentration and dependence estimates.}
\label{tab:plaintiff_infrastructure}
\small
\begin{tabular}{p{0.42\linewidth}p{0.20\linewidth}p{0.15\linewidth}p{0.15\linewidth}}
\toprule
Finding & Estimate & Inference & Interpretation \\
\midrule
Plaintiff-attorney vs. plaintiff top-10 share, 1983--2022 & 0.822 vs. 0.148 & descriptive & concentration lies in plaintiff-side counsel \\
Same-universe plaintiff top-10 share & 0.207 & descriptive & denominator alignment does not close the gap \\
Large plaintiffs' mean top-1 attorney share & 0.783 & descriptive & high-volume plaintiffs rely on dominant counsel \\
Multi-attorney top-10 share range & 0.813--0.824 & descriptive & concentration is robust to multi-attorney handling \\
Top-10 bridge, no volume control: extra plaintiffs / addresses & 297.8 / 532.5 & both \(p<0.001\) & leading attorneys broker many clients and places at scale \\

Top-10 bridge, log-volume control: extra plaintiffs / addresses & 210.9 / 377.7 & both \(p<0.001\) & residual scale effect survives log-volume adjustment \\
Shared-counsel distance: same minus different attorney & -0.129 & \(p<0.001\) & plaintiffs sharing counsel file more similarly \\
\bottomrule
\end{tabular}
\end{table}

\subsection{Repeat-address filing and tenant recurrence}\label{sec:results_repeat}

Repeated filing is pervasive. From 1969--2022, 48.8\% of residential filings with a cleaned address occur at addresses with at least one prior filing in the preceding year, and 23.6\% occur at addresses with six or more prior filings (\(N=743{,}448\) residential cases with cleaned address keys; ZIP/census fallback is excluded). Procedural outcomes shift with repetition: default rises from 46.1\% at churn 0 to 58.0\% at churn 6+, while judgment by agreement falls from 32.3\% to 23.1\%. Repeat-address cases are processed through a more default-heavy and less agreement-heavy pathway.

\begin{figure}[H]
    \centering
    \includegraphics[width=0.66\linewidth]{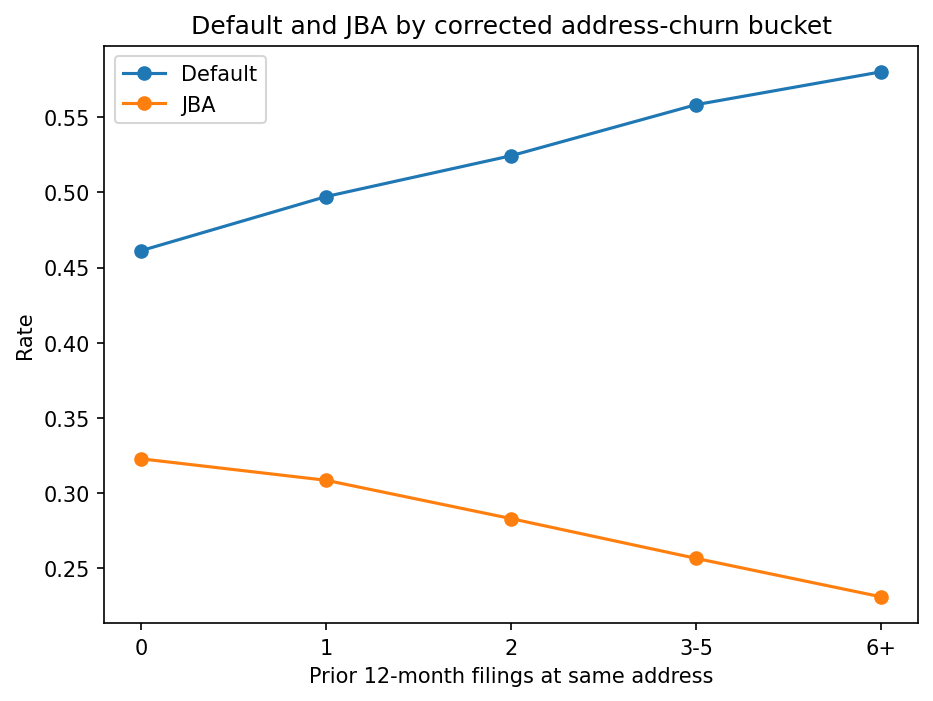}
    \caption{Default and judgment by agreement by corrected address-churn bucket. Repeated addresses become more default-heavy and less agreement-heavy as prior same-address filing accumulates.}
    \label{fig:repeated_units_churn}
\end{figure}

The pattern is not a property-scale artifact. Conditional on amount sought, representation, housing status, ZIP fixed effects, and year fixed effects, churn 6+ relative to churn 0 is associated with a 4.21 percentage-point increase in default (SE 0.44), a 9.08 percentage-point reduction in judgment by agreement (SE 0.58), and a 2.03 percentage-point reduction in served writ (SE 0.38); all coefficients are significant at \(p<0.001\) (\figref{fig:repeated_units_churn}).

Repeat addresses are dominated by same-plaintiff activity. In the repeat-address universe with plaintiff identifiers, 77.9\% of cases have a same-plaintiff prior at that address. Among exclusive repeats, 67.2\% are same-plaintiff-only rather than other-plaintiff-only (exact binomial, \(p<0.001\)). The signal is not diffuse neighborhood churn but the same plaintiff returning to the same property.

Tenant recurrence behaves differently from property recurrence. Conservative tenant-name keys yield repeat observed exposure in 31.7\% of keyed residential cases, with 16.4\% recurring within the preceding 12 months. Among repeat-tenant cases, 60.7\% reappear at a new address and 69.6\% reappear under a different plaintiff, both rejecting the 50\% benchmark at \(p<0.001\). The most common tenant key accounts for 59 cases (0.008\% of keyed cases), so the measure does not collapse the docket into a few generic aliases. These estimates capture name-based recurrence, not verified person-level recurrence.

\begin{table}[H]
\centering
\caption{Repeat-address filing and tenant recurrence: selected estimates.}
\label{tab:repeat_filing}
\small
\begin{tabular}{p{0.45\linewidth}p{0.18\linewidth}p{0.16\linewidth}p{0.14\linewidth}}
\toprule
Finding & Estimate & Inference & Interpretation \\
\midrule
Cases at addresses with any / six or more prior filings & 0.488 / 0.236 & descriptive & repeated filing is a large part of the docket \\
Default at churn 0 vs. churn 6+ & 0.461 vs. 0.580 & descriptive & repeated places are more default-heavy \\
JBA at churn 0 vs. churn 6+ & 0.323 vs. 0.231 & descriptive & repeated places are less agreement-heavy \\
Controlled churn-6+ effect on default / JBA & +0.042 / -0.091 & both \(p<0.001\) & churn effects persist with controls \\
Repeat-address cases with same-plaintiff prior & 0.779 & descriptive & repeat-address filing is usually same-plaintiff \\
Same-plaintiff-only among exclusive repeat-address cases & 0.672 & \(p<0.001\) & same-plaintiff repeats dominate other-plaintiff repeats \\
Repeat tenant ever / within 12 months & 0.317 / 0.164 & descriptive & repeated observed tenant-name exposure is common \\
Repeat tenants at new address / different plaintiff & 0.607 / 0.696 & both \(p<0.001\) & recurrence is system-level, not a single dyad \\
\bottomrule
\end{tabular}
\end{table}

\subsection{Specialist-counsel adoption}\label{sec:results_specialist}

What changes when a plaintiff switches to a top-10 attorney? Strict switching is much narrower than the loose ``first top-10 case'' definition: of 61{,}769 cases meeting the loose criterion, only 2{,}091 reflect an actual attorney change (one per plaintiff). The remaining 59{,}678 are existing-attorney threshold crossings, in which the same attorney moves into the top 10 without any change to the plaintiff's representation.

Filing counts shift around the strict switch. The CSDID monthly estimator compares each strict-switch cohort with never-treated or not-yet-treated plaintiffs. At the switch month the filing-count ATT is +1.45 (SE 0.083, \(p<0.001\)), partly mechanical because the switch itself is observed through a filing. Post-switch months 1--4 remain positive and significant (+0.174, +0.192, +0.153, +0.096; all \(p<0.05\)). A positive pre-switch coefficient at month \(-2\) (+0.056, \(p=0.006\)) and rejection of flat leads under plaintiff-cluster influence-function inference indicate that the switch coincides with broader filing reorganization. We report these estimates as identification diagnostics, not as evidence of exogenous attorney assignment.

Case-level outcomes show a stable post-switch pattern. Within plaintiff and month, strict top-10 exposure is associated with a 4.82 percentage-point reduction in the probability of judgment by agreement (SE 0.76, \(p<0.001\)), a 2.48 percentage-point increase in the probability of a served writ of possession (SE 0.87, \(p=0.004\)), and a 0.60 percentage-point reduction in the fee-to-amount-sought share (SE 0.22, \(p=0.006\)). The plaintiff--property design yields the same pattern with comparable magnitudes: judgment by agreement falls by 3.53 points (\(p<0.001\)), served writ rises by 2.86 points (\(p=0.007\)), and fee share falls by 0.82 points (\(p<0.001\)).

The largest changes are in agreement language. Within plaintiff and month, strict top-10 exposure raises deadline language by 10.21 percentage points (\(p<0.001\)) and move-out language by 2.91 percentage points (\(p=0.010\)), and lowers waiver language by 10.85 percentage points (\(p=0.002\)) and lockout-trigger language by 5.33 percentage points (\(p=0.030\)). The lockout-trigger estimate is smaller than in earlier drafts because the text parser no longer matches the bare token ``eviction.'' Deadlines and move-out clauses rise while waivers, lockout-triggers, fee share, and judgment-by-agreement rates fall---a reorganization of contractual style rather than a uniform tightening.

A placebo using existing-attorney threshold crossings calibrates the interpretation: when an existing attorney crosses the top-10 threshold without an attorney change, judgment by agreement falls by 3.26 percentage points (SE 1.42, \(p=0.022\)). Roughly two-thirds of the strict-switch JBA effect (\(4.82\) pp) is therefore not unique to attorney-change events. We interpret the strict-switch coefficients as within-plaintiff mechanism evidence, not as a clean causal effect of attorney assignment.

\begin{figure}[H]
    \centering
    \includegraphics[width=0.70\linewidth]{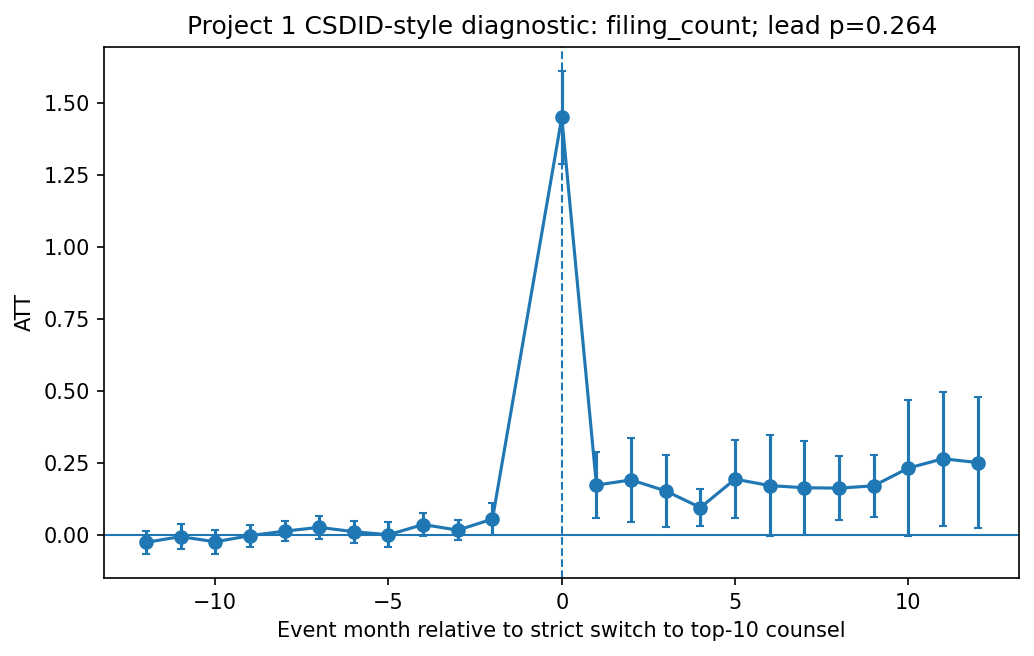}
    \caption{CSDID filing-count diagnostic around strict switching into lagged top-10 counsel. Each strict-switch cohort is compared with never-treated or not-yet-treated plaintiffs; the estimates are used as timing diagnostics rather than as stand-alone causal effects.}
    \label{fig:csdid_filing_count}
\end{figure}

\begin{table}[H]
\centering
\caption{Strict top-10 counsel switching: filing-count diagnostics and within-unit case outcomes.}
\label{tab:specialist}
\small
\begin{tabular}{p{0.42\linewidth}p{0.20\linewidth}p{0.16\linewidth}p{0.15\linewidth}}
\toprule
Finding & Estimate & Inference & Interpretation \\
\midrule
Strict switch cases / plaintiffs & 2{,}091 / 2{,}091 & descriptive & true attorney changes into top-10 counsel \\
Loose first-top-10 vs. existing-attorney crossings & 61{,}769 vs. 59{,}678 & descriptive & most loose adoption is threshold drift \\
CSDID filing count, event month & +1.449 & \(p<0.001\) & filing reorganization at switch month \\
CSDID filing count, months +1 to +4 & +0.096 to +0.192 & all \(p<0.05\) & filings remain elevated post-switch \\
Within-plaintiff JBA & -0.0482 & \(p<0.001\) & fewer judgments by agreement \\
Within-plaintiff served writ & +0.0248 & \(p=0.004\) & more served writ \\
Within-plaintiff fee share & -0.0060 & \(p=0.006\) & lower fee/cost share \\
Within-plaintiff lockout-trigger language & -0.0533 & \(p=0.030\) & less corrected lockout-trigger language \\
Within-plaintiff deadline / waiver language & +0.102 / -0.109 & both \(p<0.01\) & agreement language is reorganized \\
Existing-attorney threshold placebo, JBA & -0.0326 & \(p=0.022\) & warns against clean causal language \\
\bottomrule
\end{tabular}
\end{table}

\subsection{Exploratory attorney-style convergence}\label{sec:results_convergence}

Do strict switchers move toward the style profile of the top-10 attorney they hire? Style is defined by the pre-switch case-mix profile of the attorney's other clients, excluding the focal plaintiff. Under the expanded-window specification, 105 switchers and 30 matched placebos meet the case-count and attorney-style coverage requirements. The Mahalanobis switcher-minus-placebo convergence difference is +1.213 (95\% CI [0.139, 2.571], \(p=0.022\)): switchers move closer to the hired attorney's existing-client profile than matched placebos do. Because attorney choice is not exogenous and the placebo sample is small, we treat this as exploratory mechanism evidence, not as evidence that attorneys impose style on clients.

\begin{figure}[H]
    \centering
    \includegraphics[width=0.66\linewidth]{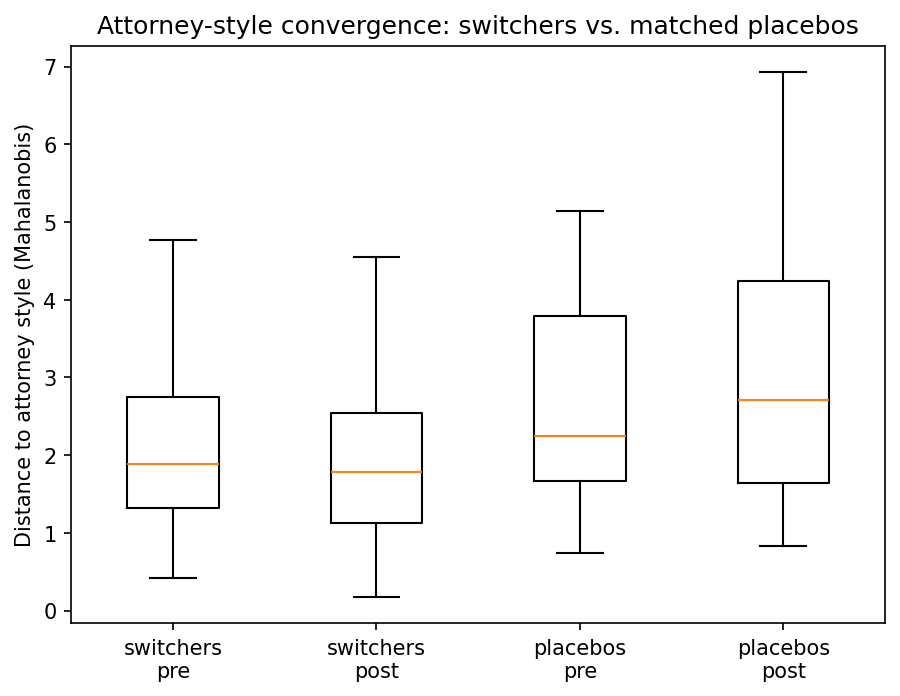}
\caption{Exploratory attorney-style convergence. The Mahalanobis switcher-minus-placebo contrast is positive and significant at \(p=0.022\); the Euclidean contrast is positive but marginal (\(p=0.062\); SI Table~\ref{tab:si_style_convergence}). The evidence is interpreted as exploratory mechanism evidence, not causal proof.}
    \label{fig:style_convergence}
\end{figure}

\section{Discussion, conclusion, and limitations}\label{sec:discussion}

\subsection{Principal contribution}

Eviction in Philadelphia is organized by upstream legal infrastructure. Filings are routed through a concentrated plaintiff-side bar; plaintiffs depend heavily on a small number of recurring attorneys; the same addresses recur in the docket; and tenant names reappear under changing address and plaintiff conditions. Strict switches into specialist plaintiff-side counsel coincide with filing reorganization and with within-unit shifts in bargaining and agreement language. Throughout, the analytic object is the filing-side pipeline that delivers cases into court, prior to bargaining, adjudication, or enforcement.

\subsection{Policy implications}

Interventions targeted at individual landlords or individual cases miss a key institutional layer. Plaintiff-side legal intermediaries concentrate court access, carry recurring client relationships, and span many properties; policy monitoring should therefore be place- and plaintiff-sensitive, not only tenant-sensitive. Because specialist-counsel adoption is associated with filing reorganization and with within-unit shifts in agreement language, the structure of the plaintiff-side legal-services market is itself a policy variable.

\subsection{Epistemic scope and limitations}

We do not claim that plaintiff-side concentration alone causes eviction, nor that every reported association admits a quasi-experimental interpretation. Concentration, plaintiff dependence, bridge structure, repeat-address filing, and tenant recurrence are descriptive institutional facts. The specialist-counsel module is more identifying but remains bounded: filing-margin estimates show reorganization around switching rather than a clean shock, while within-plaintiff and within-plaintiff--property estimates provide the most credible evidence on conditional case processing. The CSDID design improves on staggered two-way fixed effects by restricting comparisons to never-treated and not-yet-treated plaintiffs, but it does not make attorney choice exogenous.

Several measurement caveats apply. Tenant recurrence is name-based and should not be read as verified person-level recurrence. Plaintiff entity resolution is conservative and does not merge fuzzy-matched entities without a verified manual map. The bridge-attorney result is specification-sensitive, particularly in its relationship to top-client concentration. Style convergence is statistically significant in the Mahalanobis metric but exploratory: the matched-placebo sample is small (\(N=30\)) and attorneys are not randomly assigned to plaintiffs.

\subsection{Conclusion}

Eviction in Philadelphia is not a sum of independent landlord--tenant disputes. It is a legal infrastructure organized through concentrated plaintiff-side counsel, recurring plaintiffs, repeated addresses, and tenants who circulate through the docket. Characterizing this upstream layer is a precondition for analyzing how individual cases are bargained, written, monetized, and enforced.

\section*{Acknowledgments}
The authors thank Philadelphia Legal Assistance and the maintainers of the public docket data, as well as the broader civil-justice research community, for methodological guidance on the use of eviction records.

\section*{Data and Code Availability}
The Philadelphia Municipal Court dataset is available from Philadelphia Legal Assistance at \url{https://docs.philalegal.org/index.php/s/w9IQZrb8eDqXJkU}. Code to clean the data, construct measures, and reproduce the tables and figures is available at \url{https://github.com/MariosPapamix}.

\bibliographystyle{apalike}
\bibliography{Eviction}

@article{hartman2003evictions,
  title={Evictions: The hidden housing problem},
  author={Hartman, Chester and Robinson, David},
  journal={Housing Policy Debate},
  volume={14},
  number={4},
  pages={461--501},
  year={2003},
  publisher={Taylor \& Francis}
}

@article{de2020two,
  title={Two-way fixed effects estimators with heterogeneous treatment effects},
  author={De Chaisemartin, Cl{\'e}ment and d’Haultfoeuille, Xavier},
  journal={American economic review},
  volume={110},
  number={9},
  pages={2964--2996},
  year={2020},
  publisher={American Economic Association 2014 Broadway, Suite 305, Nashville, TN 37203}
}

@article{sun2021estimating,
  title={Estimating dynamic treatment effects in event studies with heterogeneous treatment effects},
  author={Sun, Liyang and Abraham, Sarah},
  journal={Journal of econometrics},
  volume={225},
  number={2},
  pages={175--199},
  year={2021},
  publisher={Elsevier}
}

@article{goodman2021difference,
  title={Difference-in-differences with variation in treatment timing},
  author={Goodman-Bacon, Andrew},
  journal={Journal of econometrics},
  volume={225},
  number={2},
  pages={254--277},
  year={2021},
  publisher={Elsevier}
}

@article{callaway2021difference,
  title={Difference-in-differences with multiple time periods},
  author={Callaway, Brantly and Sant’Anna, Pedro HC},
  journal={Journal of econometrics},
  volume={225},
  number={2},
  pages={200--230},
  year={2021},
  publisher={Elsevier}
}

@article{desmond2012eviction,
  title={Eviction and the reproduction of urban poverty},
  author={Desmond, Matthew},
  journal={American journal of sociology},
  volume={118},
  number={1},
  pages={88--133},
  year={2012},
  publisher={University of Chicago Press Chicago, IL}
}

@article{desmond2015eviction,
  title={Eviction's fallout: housing, hardship, and health},
  author={Desmond, Matthew and Kimbro, Rachel Tolbert},
  journal={Social forces},
  volume={94},
  number={1},
  pages={295--324},
  year={2015},
  publisher={Oxford University Press}
}

@article{gromis2022estimating,
  title={Estimating eviction prevalence across the United States},
  author={Gromis, Ashley and Fellows, Ian and Hendrickson, James R and Edmonds, Lavar and Leung, Lillian and Porton, Adam and Desmond, Matthew},
  journal={Proceedings of the National Academy of Sciences},
  volume={119},
  number={21},
  pages={e2116169119},
  year={2022},
  publisher={National Academy of Sciences}
}

@article{graetz2023comprehensive,
  title={A comprehensive demographic profile of the US evicted population},
  author={Graetz, Nick and Gershenson, Carl and Hepburn, Peter and Porter, Sonya R and Sandler, Danielle H and Desmond, Matthew},
  journal={Proceedings of the National Academy of Sciences},
  volume={120},
  number={41},
  pages={e2305860120},
  year={2023},
  publisher={National Academy of Sciences}
}

@article{collinson2024eviction,
  title={Eviction and poverty in American cities},
  author={Collinson, Robert and Humphries, John Eric and Mader, Nicholas and Reed, Davin and Tannenbaum, Daniel and Van Dijk, Winnie},
  journal={The Quarterly Journal of Economics},
  volume={139},
  number={1},
  pages={57--120},
  year={2024},
  publisher={Oxford University Press}
}

@article{nelson2021evictions,
  title={Evictions: The comparative analysis problem},
  author={Nelson, Kyle and Garboden, Philip and McCabe, Brian J and Rosen, Eva},
  journal={Housing Policy Debate},
  volume={31},
  number={3-5},
  pages={696--716},
  year={2021},
  publisher={Taylor \& Francis}
}

@article{porton2021inaccuracies,
  title={Inaccuracies in eviction records: Implications for renters and researchers},
  author={Porton, Adam and Gromis, Ashley and Desmond, Matthew},
  journal={Housing Policy Debate},
  volume={31},
  number={3-5},
  pages={377--394},
  year={2021},
  publisher={Taylor \& Francis}
}

@article{summers2025pathways,
  title={Pathways to eviction},
  author={Summers, Nicole and Steil, Justin},
  journal={Law \& Social Inquiry},
  volume={50},
  number={1},
  pages={129--169},
  year={2025},
  publisher={Cambridge University Press}
}

@article{galanter1974haves,
  title={Why the “haves” come out ahead: Speculations on the limits of legal change},
  author={Galanter, Marc},
  journal={Law \& society review},
  volume={9},
  number={1},
  pages={95--160},
  year={1974},
  publisher={Cambridge University Press \& Assessment}
}

@article{sudeall2021praxis,
  title={Praxis and paradox: Inside the Black Box of eviction court},
  author={Sudeall, Lauren and Pasciuti, Daniel},
  journal={Vand. L. Rev.},
  volume={74},
  pages={1365},
  year={2021},
  publisher={HeinOnline}
}

@article{engler2010connecting,
  title={Connecting self-representation to civil Gideon: What existing data reveal about when counsel is most needed},
  author={Engler, Russell},
  journal={Fordham Urb. LJ},
  volume={37},
  pages={37},
  year={2010},
  publisher={HeinOnline}
}

@article{sabbeth2022eviction,
  title={Eviction courts},
  author={Sabbeth, Kathryn A},
  journal={U. St. Thomas LJ},
  volume={18},
  pages={359},
  year={2022},
  publisher={HeinOnline}
}

@article{wilf2021assembly,
  title={Assembly-Line Plaintiffs},
  author={Wilf-Townsend, Daniel},
  journal={Harv. L. Rev.},
  volume={135},
  pages={1704},
  year={2021},
  publisher={HeinOnline}
}

@article{aizman2025shadow,
  title={Shadow players of the eviction crisis: identifying and characterizing professional evicting attorneys in Massachusetts},
  author={Aizman, Asya and Huntley, Eric Robsky},
  journal={Housing Studies},
  pages={1--24},
  year={2025},
  publisher={Taylor \& Francis}
}

@article{a2023longer,
  title={Longer trips to court cause evictions},
  author={A. Hoffman, David and Strezhnev, Anton},
  journal={Proceedings of the National Academy of Sciences},
  volume={120},
  number={2},
  pages={e2210467120},
  year={2023},
  publisher={National Academy of Sciences}
}

@article{summers2024evicted,
  title={Evicted by Default},
  author={Summers, Nicole and Steil, Justin},
  journal={Conn. L. Rev.},
  volume={57},
  pages={1233},
  year={2024},
  publisher={HeinOnline}
}

@article{cassidy2023effects,
  title={The effects of legal representation on tenant outcomes in housing court: Evidence from New York City’s Universal Access program},
  author={Cassidy, Mike and Currie, Janet},
  journal={Journal of Public Economics},
  volume={222},
  pages={104844},
  year={2023},
  publisher={Elsevier}
}

@techreport{humphries2019does,
  title={Does eviction cause poverty? Quasi-experimental evidence from Cook County, IL},
  author={Humphries, John Eric and Mader, Nicholas S and Tannenbaum, Daniel I and Van Dijk, Winnie L},
  year={2019},
  institution={National Bureau of Economic Research}
}

@article{garboden2019serial,
  title={Serial filing: How landlords use the threat of eviction},
  author={Garboden, Philip ME and Rosen, Eva},
  journal={City \& Community},
  volume={18},
  number={2},
  pages={638--661},
  year={2019},
  publisher={SAGE Publications Sage CA: Los Angeles, CA}
}

@article{leung2021serial,
  title={Serial eviction filing: Civil courts, property management, and the threat of displacement},
  author={Leung, Lillian and Hepburn, Peter and Desmond, Matthew},
  journal={Social Forces},
  volume={100},
  number={1},
  pages={316--344},
  year={2021},
  publisher={Oxford University Press}
}

@article{immergluck2020evictions,
  title={Evictions, large owners, and serial filings: Findings from Atlanta},
  author={Immergluck, Dan and Ernsthausen, Jeff and Earl, Stephanie and Powell, Allison},
  journal={Housing Studies},
  volume={35},
  number={5},
  pages={903--924},
  year={2020},
  publisher={Taylor \& Francis}
}

@article{hepburn2020racial,
  title={Racial and gender disparities among evicted Americans},
  author={Hepburn, Peter and Louis, Renee and Desmond, Matthew},
  journal={Sociological Science},
  volume={7},
  pages={649--662},
  year={2020}
}

@article{gomory2023racially,
  title={The racially disparate influence of filing fees on eviction rates},
  author={Gomory, Henry and Massey, Douglas S and Hendrickson, James R and Desmond, Matthew},
  journal={Housing Policy Debate},
  volume={33},
  number={6},
  pages={1463--1483},
  year={2023},
  publisher={Taylor \& Francis}
}

@article{ajayi2026landlord,
  title={Landlord responsiveness to eviction filing fees: evidence from northern New England},
  author={Ajayi, Oluwafisayo and Hobbs, Kelsi G and Wibabara, Eliane},
  journal={Regional Studies, Regional Science},
  volume={13},
  number={1},
  pages={2620937},
  year={2026},
  publisher={Taylor \& Francis}
}

@article{brito2022racial,
  title={Racial capitalism in the civil courts},
  author={Brito, Tonya L and Sabbeth, Kathryn A and Steinberg, Jessica K and Sudeall, Lauren},
  journal={Colum. L. Rev.},
  volume={122},
  pages={1243},
  year={2022},
  publisher={HeinOnline}
}

@article{rutan2021concentrated,
  title={The concentrated geography of eviction},
  author={Rutan, Devin Q and Desmond, Matthew},
  journal={The ANNALS of the American Academy of Political and Social Science},
  volume={693},
  number={1},
  pages={64--81},
  year={2021},
  publisher={SAGE Publications Sage CA: Los Angeles, CA}
}

@article{summers2023civil,
  title={Civil probation},
  author={Summers, Nicole},
  journal={Stan. L. Rev.},
  volume={75},
  pages={847},
  year={2023},
  publisher={HeinOnline}
}

@article{summers2026settlements,
  title={Settlements of Adhesion},
  author={Summers, Nicole},
  journal={University of Chicago Law Review},
  volume={93},
  number={1},
  year={2026}
}

@article{kleysteuber2006tenant,
  title={Tenant screening thirty years later: A statutory proposal to protect public records},
  author={Kleysteuber, Rudy},
  journal={Yale LJ},
  volume={116},
  pages={1344},
  year={2006},
  publisher={HeinOnline}
}

@article{reosti2020we,
  title={“We go totally subjective”: Discretion, discrimination, and tenant screening in a landlord’s market},
  author={Reosti, Anna},
  journal={Law \& Social Inquiry},
  volume={45},
  number={3},
  pages={618--657},
  year={2020},
  publisher={Cambridge University Press}
}

@article{eisenberg2024record,
  title={Record Costs: Collateral Consequences of Eviction Court Filings in Pennsylvania},
  author={Eisenberg, Alexa and Brantley, Kate},
  journal={Ann Arbor: University of Michigan, Housing Solutions for Health Equity},
  year={2024}
}

@article{brantley2025record,
  title={Record costs: examining the impact of eviction filings for tenants and their families},
  author={Brantley, Kate and Eisenberg, Alexa and Mehdipanah, Roshanak},
  journal={Housing Studies},
  pages={1--27},
  year={2025},
  publisher={Taylor \& Francis}
}

@article{hepburn2023protecting,
  title={Protecting the most vulnerable: policy response and eviction filing patterns during the COVID-19 pandemic},
  author={Hepburn, Peter and Haas, Jacob and Graetz, Nick and Louis, Renee and Rutan, Devin Q and Alexander, Anne Kat and Rangel, Jasmine and Jin, Olivia and Benfer, Emily and Desmond, Matthew},
  journal={RSF: The Russell Sage Foundation Journal of the Social Sciences},
  volume={9},
  number={3},
  pages={186--207},
  year={2023},
  publisher={RSF: The Russell Sage Foundation Journal of the Social Sciences}
}

@article{benfer2023covid,
  title={COVID-19 housing policy: State and federal eviction moratoria and supportive measures in the United States during the pandemic},
  author={Benfer, Emily A and Koehler, Robert and Mark, Alyx and Nazzaro, Valerie and Alexander, Anne Kat and Hepburn, Peter and Keene, Danya E and Desmond, Matthew},
  journal={Housing Policy Debate},
  volume={33},
  number={6},
  pages={1390--1414},
  year={2023},
  publisher={Taylor \& Francis}
}

@article{zainulbhai2022informal,
  title={Informal evictions: Measuring displacement outside the courtroom},
  author={Zainulbhai, Sabiha and Daly, Nora},
  year={2022},
  publisher={< bound method Organization. get\_name\_with\_acronym of< Organization: New~…}
}

@article{shanahan2022judges,
  title={Judges in Lawyerless Courts},
  author={Shanahan, Colleen F and Carpenter, Anna E and Steinberg, Jessica and Mark, Alyx},
  journal={Georgetown Law Journal},
  pages={509},
  year={2022}
}

@book{desmond2017evicted,
  title={Evicted: Poverty and profit in the American city},
  author={Desmond, Matthew},
  year={2017},
  publisher={Crown}
}

@article{desmond2017gets,
  title={Who gets evicted? Assessing individual, neighborhood, and network factors},
  author={Desmond, Matthew and Gershenson, Carl},
  journal={Social science research},
  volume={62},
  pages={362--377},
  year={2017},
  publisher={Elsevier}
}

@article{greenberg2016discrimination,
  title={Discrimination in evictions: empirical evidence and legal challenges},
  author={Greenberg, Deena and Gershenson, Carl and Desmond, Matthew},
  journal={Harv. CR-CLL Rev.},
  volume={51},
  pages={115},
  year={2016},
  publisher={HeinOnline}
}

@article{desmond2015forced,
  title={Forced relocation and residential instability among urban renters},
  author={Desmond, Matthew and Gershenson, Carl and Kiviat, Barbara},
  journal={Social Service Review},
  volume={89},
  number={2},
  pages={227--262},
  year={2015},
  publisher={University of Chicago Press Chicago, IL}
}

@incollection{felstiner2017emergence,
  title={The emergence and transformation of disputes: Naming, blaming, claiming…},
  author={Felstiner, William LF and Abel, Richard L and Sarat, Austin},
  booktitle={Theoretical and Empirical Studies of Rights},
  pages={255--306},
  year={2017},
  publisher={Routledge}
}

@article{mnookin1978bargaining,
  title={Bargaining in the shadow of the law: The case of divorce},
  author={Mnookin, Robert H and Kornhauser, Lewis},
  journal={Yale lJ},
  volume={88},
  pages={950},
  year={1978},
  publisher={HeinOnline}
}

@article{priest1984selection,
  title={The selection of disputes for litigation},
  author={Priest, George L and Klein, Benjamin},
  journal={The journal of legal studies},
  volume={13},
  number={1},
  pages={1--55},
  year={1984},
  publisher={The University of Chicago Law School}
}

@article{sandefur2015elements,
  title={Elements of professional expertise: Understanding relational and substantive expertise through lawyers’ impact},
  author={Sandefur, Rebecca L},
  journal={American Sociological Review},
  volume={80},
  number={5},
  pages={909--933},
  year={2015},
  publisher={Sage Publications Sage CA: Los Angeles, CA}
}

@article{pistor2019code,
  title={The code of capital: How the law creates wealth and inequality},
  author={Pistor, Katharina},
  year={2019},
  publisher={Princeton University Press}
}

@article{rabin1983revolution,
  title={Revolution in residential landlord-tenant law: causes and consequences},
  author={Rabin, Edward H},
  journal={Cornell L. Rev.},
  volume={69},
  pages={517},
  year={1983},
  publisher={HeinOnline}
}

@article{seron2001impact,
  title={The impact of legal counsel on outcomes for poor tenants in New York City's housing court: results of a randomized experiment},
  author={Seron, Carroll and Van Ryzin, Gregg and Frankel, Martin},
  journal={Law \& Society Review},
  volume={35},
  number={2},
  pages={419--434},
  year={2001},
  publisher={Cambridge University Press \& Assessment}
}

@article{bezdek1991silence,
  title={Silence in the court: Participation and subordination of poor tenants' voices in legal process},
  author={Bezdek, Barbara},
  journal={Hofstra L. Rev.},
  volume={20},
  pages={533},
  year={1991},
  publisher={HeinOnline}
}

@article{ellen2021lawyers,
  title={Do lawyers matter? Early evidence on eviction patterns after the rollout of universal access to counsel in New York City},
  author={Ellen, Ingrid Gould and O’Regan, Katherine and House, Sophia and Brenner, Ryan},
  journal={Housing Policy Debate},
  volume={31},
  number={3-5},
  pages={540--561},
  year={2021},
  publisher={Taylor \& Francis}
}

@article{summers2022eviction,
  title={Eviction court displacement rates},
  author={Summers, Nicole},
  journal={Nw. UL REv.},
  volume={117},
  pages={287},
  year={2022},
  publisher={HeinOnline}
}

@article{harris2010drawing,
  title={Drawing blood from stones: Legal debt and social inequality in the contemporary United States},
  author={Harris, Alexes and Evans, Heather and Beckett, Katherine},
  journal={American journal of sociology},
  volume={115},
  number={6},
  pages={1753--1799},
  year={2010},
  publisher={The University of Chicago Press}
}

@article{holland2011one,
  title={The one hundred billion dollar problem in small claims court: Robo-signing and lack of proof in debt buyer cases},
  author={Holland, Peter A},
  journal={J. Bus. \& Tech. L.},
  volume={6},
  pages={259},
  year={2011},
  publisher={HeinOnline}
}

@book{leibowitz2010repairing,
  title={Repairing a broken system: Protecting consumers in debt collection litigation and arbitration},
  author={Leibowitz, Jon},
  year={2010},
  publisher={DIANE Publishing}
}

\clearpage
\appendix
\counterwithin{table}{section}
\counterwithin{figure}{section}
\renewcommand{\thesection}{\Alph{section}}
\renewcommand{\thesubsection}{\thesection\arabic{subsection}}
\renewcommand{\thetable}{\thesection\arabic{table}}
\renewcommand{\thefigure}{\thesection\arabic{figure}}

\section*{Supplementary Information}\label{app:si}
\addcontentsline{toc}{section}{Supplementary Information}

This supplement reports additional robustness checks and diagnostic estimates for the filing-infrastructure analysis. The main text emphasizes corrected estimates that remove ZIP/census address fallback, exclude same-day priors, use conservative tenant keys, require strict attorney switching, and distinguish descriptive infrastructure claims from quasi-causal specialist-counsel diagnostics.

\section{Plaintiff-side legal infrastructure}\label{app:A}

\subsection{Concentration sensitivity and dominant-counsel dependence}\label{app:A1}

The denominator-alignment checks confirm that the attorney-concentration result is not an artifact of comparing represented cases to the full plaintiff universe. On the represented and named-attorney universe, the plaintiff top-10 share rises only to 0.207, compared with 0.822 for plaintiff attorneys.

\begin{table}[H]
\centering
\caption{Same-universe concentration sensitivity, 1983--2022.}
\label{tab:si_concentration_sensitivity}
\small
\begin{tabular}{p{0.55\linewidth}cc}
\toprule
Definition & Mean HHI & Mean top-10 share \\
\midrule
All residential plaintiffs & 0.0076 & 0.148 \\
Plaintiffs on represented named-attorney universe & 0.0136 & 0.207 \\
Plaintiff attorneys on represented named-attorney universe & 0.1066 & 0.822 \\
Any-listed attorney, fractional unique-case denominator & 0.1066 & 0.821 \\
Any-listed case has any top-10 attorney & -- & 0.824 \\
Case-attorney-pairing denominator & 0.1044 & 0.813 \\
\bottomrule
\end{tabular}
\end{table}

\begin{table}[H]
\centering
\caption{Dominant-counsel dependence by plaintiff volume.}
\label{tab:si_dominant_counsel}
\small
\begin{tabular}{lrrrr}
\toprule
Plaintiff volume bin & Plaintiffs & Mean top-1 attorney share & Mean unique attorneys & Mean years active \\
\midrule
2--5 & 21{,}895 & 0.855 & 1.39 & 2.02 \\
6--10 & 3{,}920 & 0.816 & 1.85 & 3.95 \\
11--25 & 2{,}569 & 0.810 & 2.10 & 5.65 \\
26--50 & 1{,}077 & 0.811 & 2.44 & 7.56 \\
51--100 & 638 & 0.790 & 2.98 & 9.98 \\
101+ & 545 & 0.783 & 4.17 & 13.46 \\
\bottomrule
\end{tabular}
\end{table}

\subsection{Bridge specifications and omnibus attorney checks}\label{app:A2}

The bridge result depends on whether the model is interpreted as a scale model or a portfolio-structure model. The scale specification asks whether top-10 attorneys span more plaintiffs and addresses than other attorneys. The volume-controlled specifications ask whether top-10 status adds portfolio breadth after conditioning on attorney-year volume.

\begin{table}[H]
\centering
\caption{Attorney-year bridge regressions under three specifications.}
\label{tab:si_bridge_specs}
\small
\begin{tabular}{p{0.28\linewidth}p{0.24\linewidth}rrr}
\toprule
Outcome & Specification & Coefficient & SE & $p$-value \\
\midrule
Unique plaintiffs & No volume control & 297.8 & 62.8 & $<0.001$ \\

Unique plaintiffs & Log total cases & 210.9 & 46.9 & $<0.001$ \\
Unique addresses & No volume control & 532.5 & 88.8 & $<0.001$ \\

Unique addresses & Log total cases & 377.7 & 69.5 & $<0.001$ \\
Top-1 plaintiff share & No volume control & -0.597 & 0.038 & $<0.001$ \\
Top-1 plaintiff share & Linear total cases & -0.508 & 0.045 & $<0.001$ \\
Top-1 plaintiff share & Log total cases & 0.637 & 0.061 & $<0.001$ \\
\bottomrule
\end{tabular}
\end{table}

Pooled attorney omnibus tests are retained as high-volume-attorney robustness checks rather than as broad attorney-population evidence. With the original controls, the min-5 threshold retains at least 108--128 attorneys across core outcomes and is the recommended broader threshold; the min-30 version retains only 26--31 attorneys and should be labeled high-volume-only.

\begin{table}[H]
\centering
\caption{Omnibus attorney-effect attrition by minimum cases per attorney.}
\label{tab:si_omnibus_attrition}
\small
\begin{tabular}{lrrrr}
\toprule
Outcome & Min cases & Complete-case attorneys & Attorneys retained & Decision \\
\midrule
Default/JBA/served writ & 5 & 875 & 128 & report with attrition warning \\
Default/JBA/served writ & 10 & 875 & 77 & report with attrition warning \\
Default/JBA/served writ & 30 & 875 & 31 & high-volume-only \\
Fee share & 5 & 755 & 108 & report with attrition warning \\
Fee share & 10 & 755 & 58 & report with attrition warning \\
Fee share & 30 & 755 & 26 & high-volume-only \\
\bottomrule
\end{tabular}
\end{table}

\section{Specialist-counsel adoption}\label{app:B}

\subsection{CSDID filing-count diagnostics}\label{app:B1}

The monthly CSDID diagnostic is the main filing-count timing diagnostic. It compares each strict-switch cohort only with never-treated or not-yet-treated plaintiffs and uses diagnostic bootstrap inference over \(ATT(g,t)\) cells. A quarterly package implementation is reported as a computational check and is not used as a basis for causal claims about the paper.

\begin{figure}[H]
    \centering
    \includegraphics[width=0.72\linewidth]{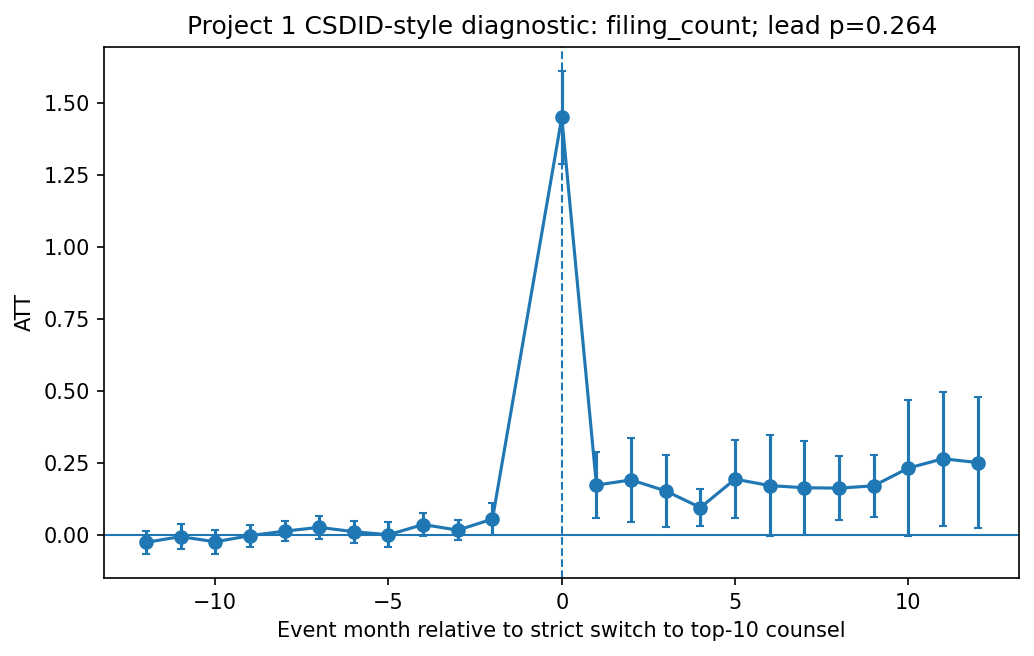}
    \caption{CSDID clean-control filing-count diagnostic for strict switches into lagged top-10 counsel.}
    \label{fig:si_csdid_style}
\end{figure}

\begin{table}[H]
\centering
\caption{CSDID monthly filing-count event-time estimates, selected significant horizons.}
\label{tab:si_csdid_monthly}
\small
\begin{tabular}{rrrrr}
\toprule
Event month & ATT & SE & 95\% CI & $p$-value \\
\midrule

-2 & 0.0559 & 0.0276 & [0.0037, 0.1070] & 0.043 \\
0 & 1.4494 & 0.0828 & [1.2941, 1.6050] & $<0.001$ \\
1 & 0.1736 & 0.0581 & [0.0836, 0.2627] & 0.0028 \\
2 & 0.1917 & 0.0748 & [0.0872, 0.2954] & 0.010 \\
3 & 0.1534 & 0.0645 & [0.0554, 0.2537] & 0.017 \\
4 & 0.0957 & 0.0323 & [0.0385, 0.1557] & 0.003 \\
\bottomrule
\end{tabular}
\end{table}

\begin{table}[H]
\centering
\caption{Quarterly software check, selected significant horizons.}
\label{tab:si_canonical_attgt}
\small
\begin{tabular}{rrrrr}
\toprule
Event quarter & ATT & SE & 95\% CI & Significant \\
\midrule

0 & 0.7757 & 0.0092 & [0.7576, 0.7938] & yes \\
1 & 0.0873 & 0.0102 & [0.0673, 0.1074] & yes \\
2 & 0.0851 & 0.0106 & [0.0644, 0.1059] & yes \\
3 & 0.0756 & 0.0106 & [0.0548, 0.0964] & yes \\
4 & 0.0661 & 0.0103 & [0.0460, 0.0863] & yes \\
\bottomrule
\end{tabular}
\end{table}

\subsection{Within-unit estimates and threshold-crossing placebo}\label{app:B2}

The within-plaintiff and within-plaintiff--property estimates are the main case-processing evidence. They compare the same plaintiff, and in the stronger design the same plaintiff--property unit, before and after strict switching.

\begin{table}[H]
\centering
\caption{Within-plaintiff and within-plaintiff--property estimates for strict top-10 counsel switching.}
\label{tab:si_specialist_case_fe}
\small
\begin{tabular}{p{0.27\linewidth}p{0.28\linewidth}p{0.28\linewidth}}
\toprule
Outcome & Within-plaintiff FE & Within-plaintiff--property FE \\
\midrule

JBA & -0.0482 ($p<0.001$) & -0.0353 ($p<0.001$) \\
Served writ & +0.0248 ($p=0.004$) & +0.0286 ($p=0.0066$) \\
Fee share & -0.0060 ($p=0.006$) & -0.0082 ($p<0.001$) \\
Move-out clause & +0.0291 ($p=0.010$) & +0.0340 ($p=0.0166$) \\
Lockout trigger & -0.0533 ($p=0.030$) & -- \\
Deadline clause & +0.1021 ($p<0.001$) & +0.1295 ($p<0.001$) \\
Waiver clause & -0.1085 ($p=0.0019$) & -0.0710 ($p=0.0337$) \\
\bottomrule
\end{tabular}
\end{table}

\begin{table}[H]
\centering
\caption{Placebo: existing attorney crosses the top-10 threshold without attorney change.}
\label{tab:si_threshold_placebo}
\small
\begin{tabular}{lrrr}
\toprule
Outcome & Coefficient & SE & $p$-value \\
\midrule
JBA & -0.0326 & 0.0142 & 0.022 \\

\bottomrule
\end{tabular}
\end{table}

\subsection{JBA strictness sensitivity}\label{app:B3}

Composite strictness scores are reported as sensitivity checks rather than headline outcomes because the components move in different directions. The first principal component is positive and significant. The individual components give the clearest substantive interpretation: deadline language rises, while waiver and lockout-trigger language fall.

\begin{table}[H]
\centering
\caption{JBA strictness sensitivity and component estimates, within-plaintiff model.}
\label{tab:si_jba_components}
\small
\begin{tabular}{lrrr}
\toprule
Outcome & Coefficient & SE & $p$-value \\
\midrule

PC1 strictness & +0.0735 & 0.0180 & $<0.001$ \\
Move-out clause & +0.0291 & 0.0114 & 0.010 \\
Lockout trigger & -0.0533 & 0.0246 & 0.030 \\
Time-is-essence clause & -0.0939 & 0.0386 & 0.015 \\
Deadline clause & +0.1021 & 0.0195 & $<0.001$ \\
Waiver clause & -0.1085 & 0.0350 & 0.0019 \\
\bottomrule
\end{tabular}
\end{table}

\section{Repeat-address filing and tenant recurrence}\label{app:C}

\subsection{Same-plaintiff repeat-address filing and churn controls}\label{app:C1}

Same-plaintiff repetition dominates repeated-address filing. In the repeat-address universe with plaintiff identifiers, 77.9\% of repeat-address cases have a same-plaintiff prior at that address. Among exclusive repeat-address cases, 67.2\% are same-plaintiff-only cases rather than other-plaintiff-only cases.

\begin{table}[H]
\centering
\caption{Same-plaintiff repeat-address decomposition.}
\label{tab:si_repeat_address_decomp}
\small
\begin{tabular}{p{0.58\linewidth}cc}
\toprule
Statistic & Estimate & $p$-value \\
\midrule
Repeat-address cases with any same-plaintiff prior & 0.779 & -- \\
Same-plaintiff-only share among exclusive repeat-address cases & 0.672 & $<0.001$ \\
Other-plaintiff-only share among exclusive repeat-address cases & 0.328 & -- \\
Both same- and other-plaintiff prior among repeat-address cases & 0.326 & -- \\
\bottomrule
\end{tabular}
\end{table}

\begin{table}[H]
\centering
\caption{Controlled churn models. Relative to churn 0, higher churn buckets remain associated with more default and less JBA after case-mix, ZIP, and year controls.}
\label{tab:si_churn_controls}
\small
\begin{tabular}{llrrr}
\toprule
Outcome & Term & Coefficient & Standard error & $p$-value \\
\midrule
Default & Churn 1 & 0.0243 & 0.0024 & $<0.001$ \\
Default & Churn 2 & 0.0325 & 0.0039 & $<0.001$ \\
Default & Churn 3--5 & 0.0441 & 0.0040 & $<0.001$ \\
Default & Churn 6+ & 0.0421 & 0.0044 & $<0.001$ \\
JBA & Churn 1 & -0.0066 & 0.0024 & 0.006 \\
JBA & Churn 2 & -0.0197 & 0.0038 & $<0.001$ \\
JBA & Churn 3--5 & -0.0574 & 0.0040 & $<0.001$ \\
JBA & Churn 6+ & -0.0908 & 0.0058 & $<0.001$ \\
Served writ & Churn 1 & 0.0116 & 0.0021 & $<0.001$ \\

Served writ & Churn 6+ & -0.0203 & 0.0038 & $<0.001$ \\
\bottomrule
\end{tabular}
\end{table}

\subsection{Tenant recurrence and system circulation}\label{app:C2}

Tenant recurrence is common but not usually confined to a single address or plaintiff. Among repeat-tenant-name cases, 60.7\% reappear at a new address and 69.6\% reappear under a different plaintiff. Repeat-tenant history also changes the composition of later cases: repeat-address exposure rises from 42.6\% for new-to-system tenant names to more than 60\% for recurrent tenant-name groups.

\begin{table}[H]
\centering
\caption{Tenant recurrence and already-exposed tenant targeting.}
\label{tab:si_tenant_recurrence}
\small
\begin{tabular}{p{0.57\linewidth}p{0.18\linewidth}p{0.12\linewidth}}
\toprule
Finding & Estimate & $p$-value \\
\midrule
Share of cases involving repeat tenant names & 0.317 & -- \\
Repeat-tenant cases with recurrence in prior 12 months & 0.164 & -- \\
Repeat-tenant reappearance at new address & 0.607 & $<0.001$ \\
Repeat-tenant reappearance under different plaintiff & 0.696 & $<0.001$ \\
Largest tenant key share of keyed cases & 0.00008 & -- \\
\bottomrule
\end{tabular}
\end{table}

\section{Attorney-style convergence}\label{app:D}

The attorney-style convergence analysis is exploratory. It compares a strict switcher's movement toward the new attorney's existing-client profile with matched never-switcher placebos. The Mahalanobis difference is statistically significant.

\begin{table}[H]
\centering
\caption{Attorney-style convergence, switchers versus matched placebos.}
\label{tab:si_style_convergence}
\small
\begin{tabular}{lrrr}
\toprule
Metric & Switcher-minus-placebo difference & 95\% CI & $p$-value \\
\midrule
Mahalanobis & 1.213 & [0.139, 2.571] & 0.022 \\
Euclidean   & 1.039 & [-0.029, 2.255] & 0.062 \\

\bottomrule
\end{tabular}
\end{table}

\end{document}